\documentclass[twocolumn,showpacs,preprintnumbers,amsmath,amssymb,floatfix]{revtex4}

\usepackage{graphicx}
\usepackage{dcolumn}
\usepackage{bm}
\usepackage{epsfig}
\usepackage{epstopdf}

\begin{document}

\title{Percolation thresholds on 2D Voronoi networks and Delaunay triangulations}

\author{Adam M. Becker}
\email{beckeram@umich.edu}
\affiliation{Department of Physics, University of Michigan, Ann Arbor MI 48109-1040}
\author{Robert M. Ziff}
\email{rziff@umich.edu}
\affiliation{Center for the Study of Complex Systems and Department of Chemical Engineering, University of Michigan, Ann Arbor MI 48109-2136}

\date{\today}
\begin{abstract}
The site percolation threshold for the random Voronoi network is determined numerically for the first time, with the result $p_c = 0.71410 \pm 0.00002$, using Monte-Carlo simulation on periodic systems of up to 40000 sites.  The result is very close to the recent theoretical estimate $p_c \approx 0.7151$ of Neher, Mecke, and Wagner.  For the bond threshold on the Voronoi network, we find $p_c = 0.666931 \pm 0.000005$, implying that for its dual, the Delaunay triangulation, $p_c = 0.333069 \pm 0.000005$.  These results rule out the conjecture by Hsu and Huang that the bond thresholds are 2/3 and 1/3 respectively, but support the conjecture of Wierman that for fully triangulated lattices other than the regular triangular lattice, the bond threshold is less than $2 \sin \pi/18 \approx 0.3473$.
\end{abstract}
\maketitle
\section{Introduction}
\begin{figure}[ t] 
  \centering
   \includegraphics[width=0.45\textwidth]{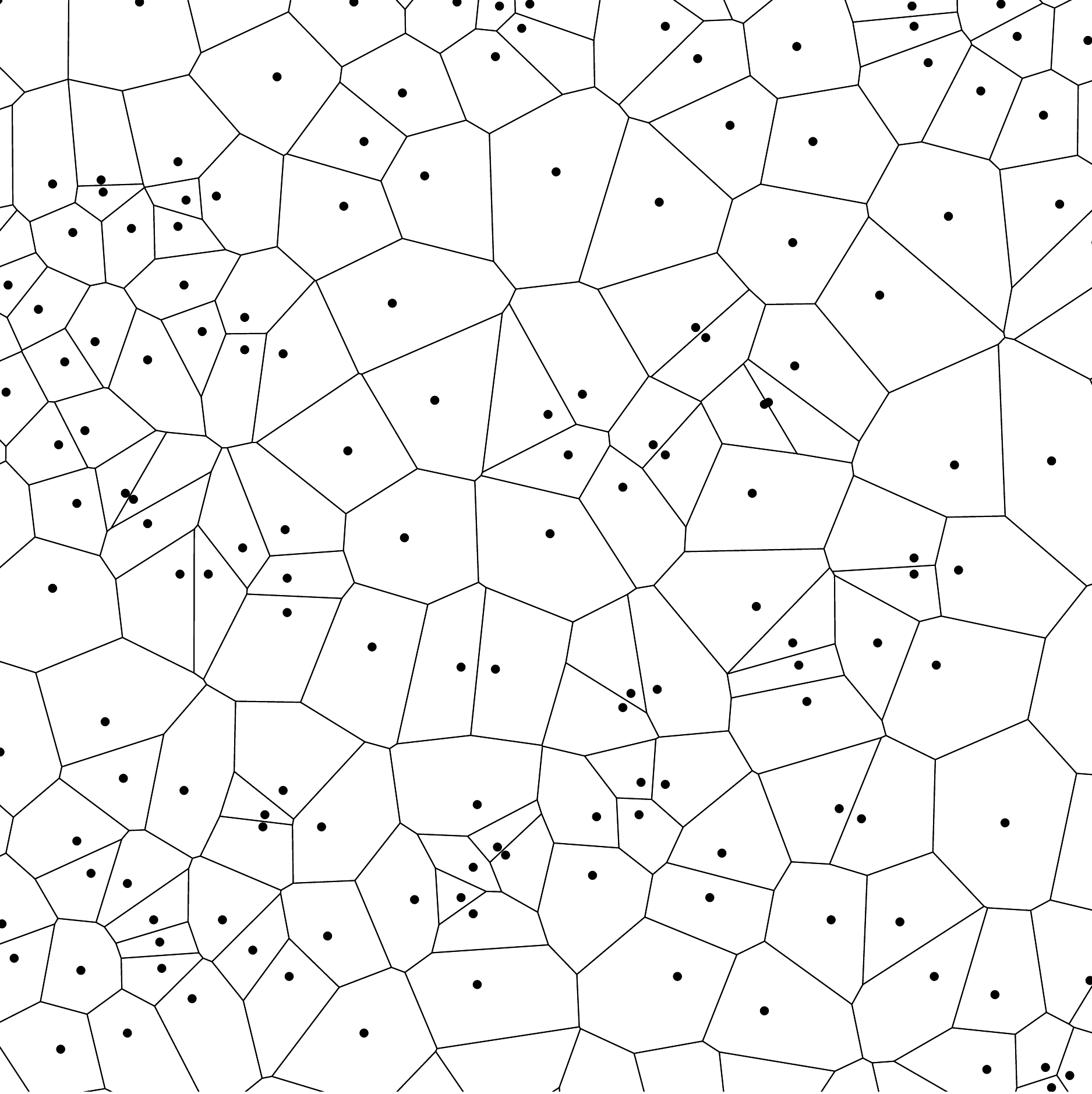} 
   \caption{Voronoi diagram with a Poisson distribution of generating points.}
   \label{fig:Voronoi}
\end{figure}

The Voronoi diagram \cite{Voronoi1908} for a given set of points on a plane (Fig.\  \ref{fig:Voronoi}) is simple to define. Given some set of points $P$ on a plane $\mathbb R^2$, the Voronoi diagram divides the plane $\mathbb R^2$ into polygons, each containing exactly one member of $P$. Each point's polygon cordons off the portion of $\mathbb R^2$ that is closer to that point than to any other member of $P$. More precisely, the Voronoi polygon around $p_i \in P$ contains all locations on $\mathbb R^2$ that are closer to $p_i$ than to any other element of $P$. The total Voronoi diagram is the set of all the Voronoi polygons for $P$ on $\mathbb R^2$; the Voronoi {\it network} is the set of vertices and edges of the Voronoi diagram.

\begin{figure}[ t] 
  \centering
   \includegraphics[width=0.45\textwidth]{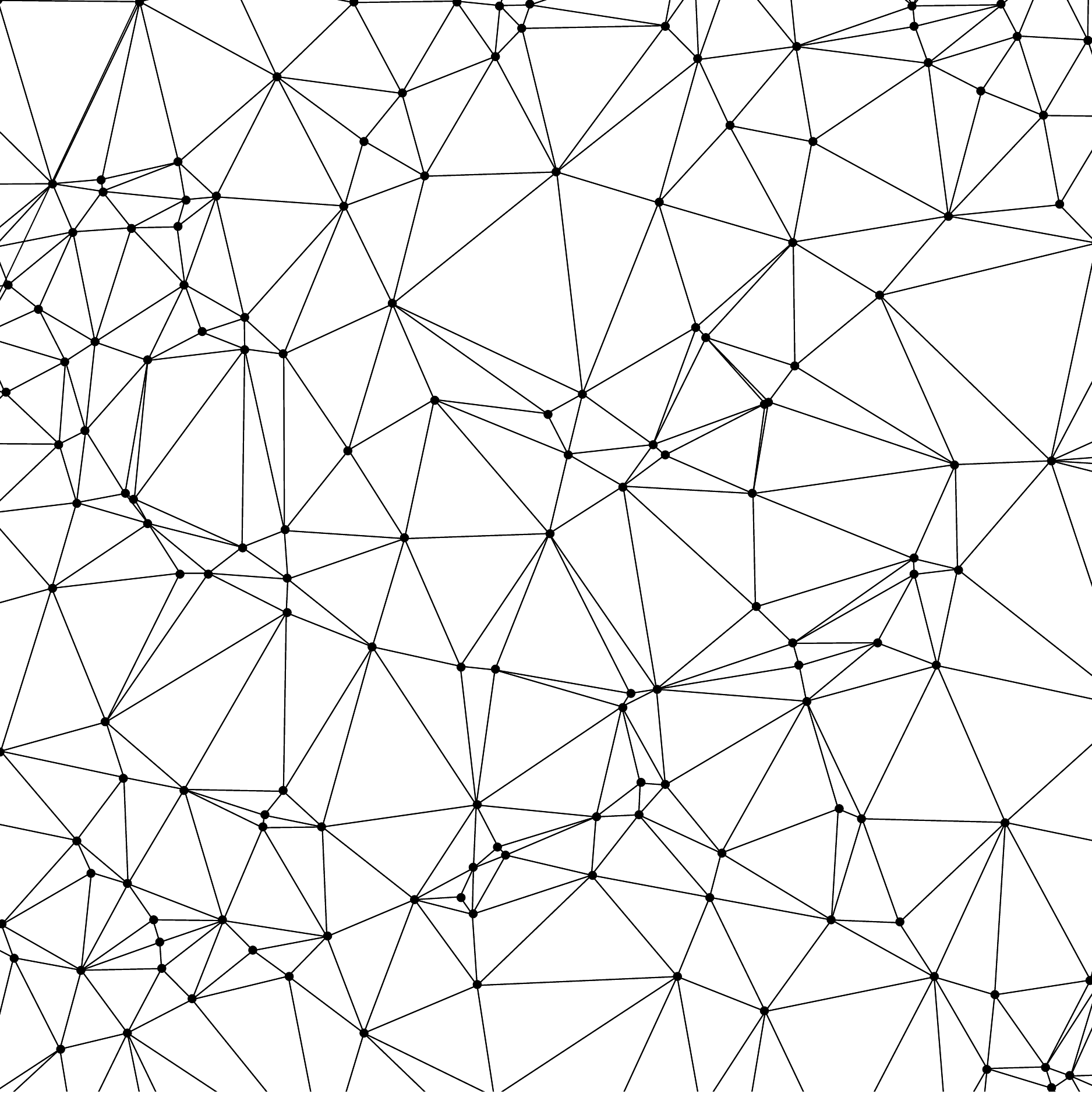} 
   \caption{Delaunay triangulation for the same set of generating points as in Fig.\  \ref{fig:Voronoi}.  The generating points become the vertices in this network.}
   \label{fig:Delaunay}
\end{figure}

\begin{figure}[ t] 
   \centering
   \includegraphics[width=0.45\textwidth]{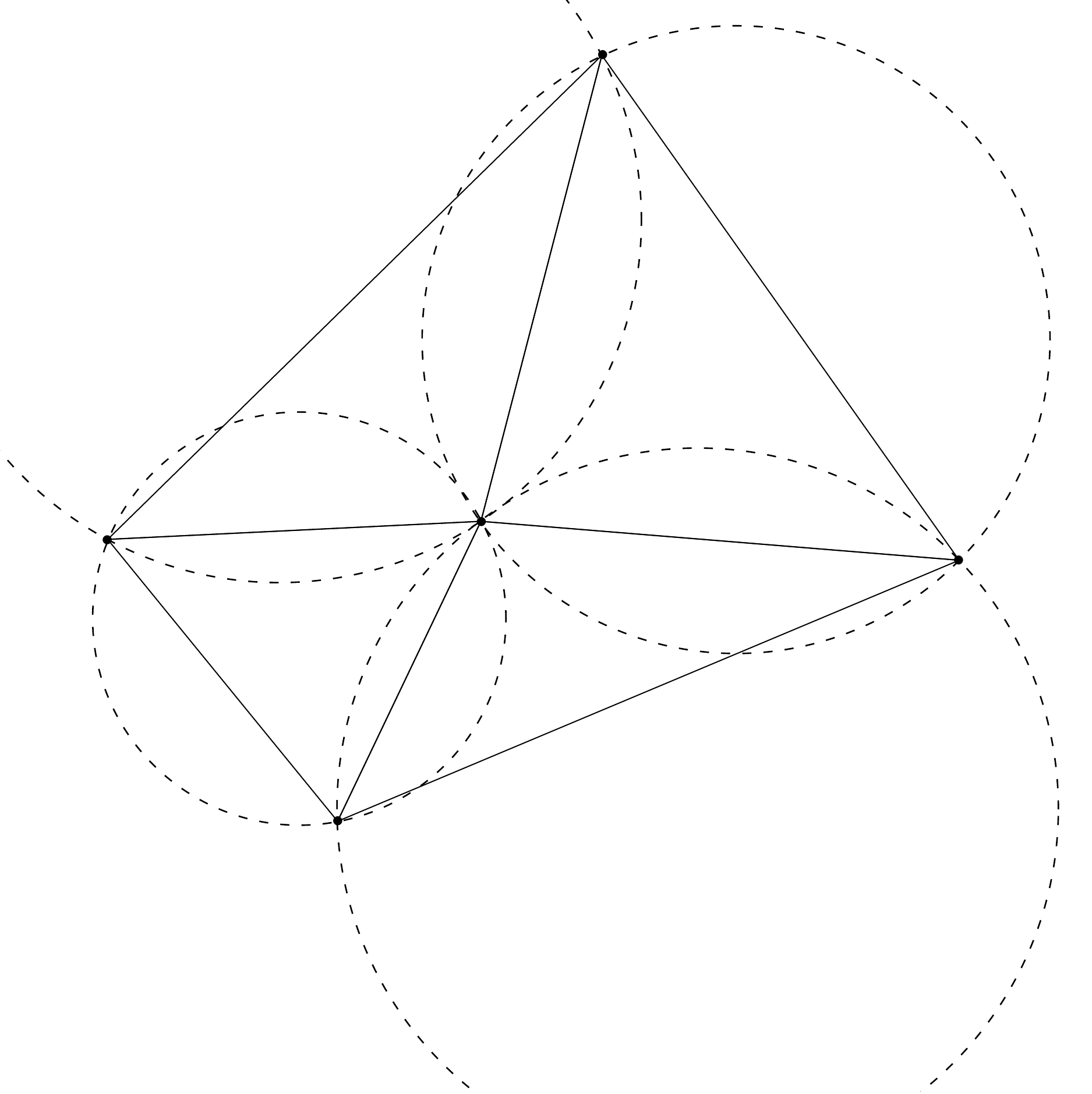} 
   \caption{The Delaunay triangulation for a set of five points, along with the associated circumcircles.}
   \label{fig:DelaunayCircumcircles}
\end{figure}

\begin{figure}[ t] 
   \centering
   \includegraphics[width=0.45\textwidth]{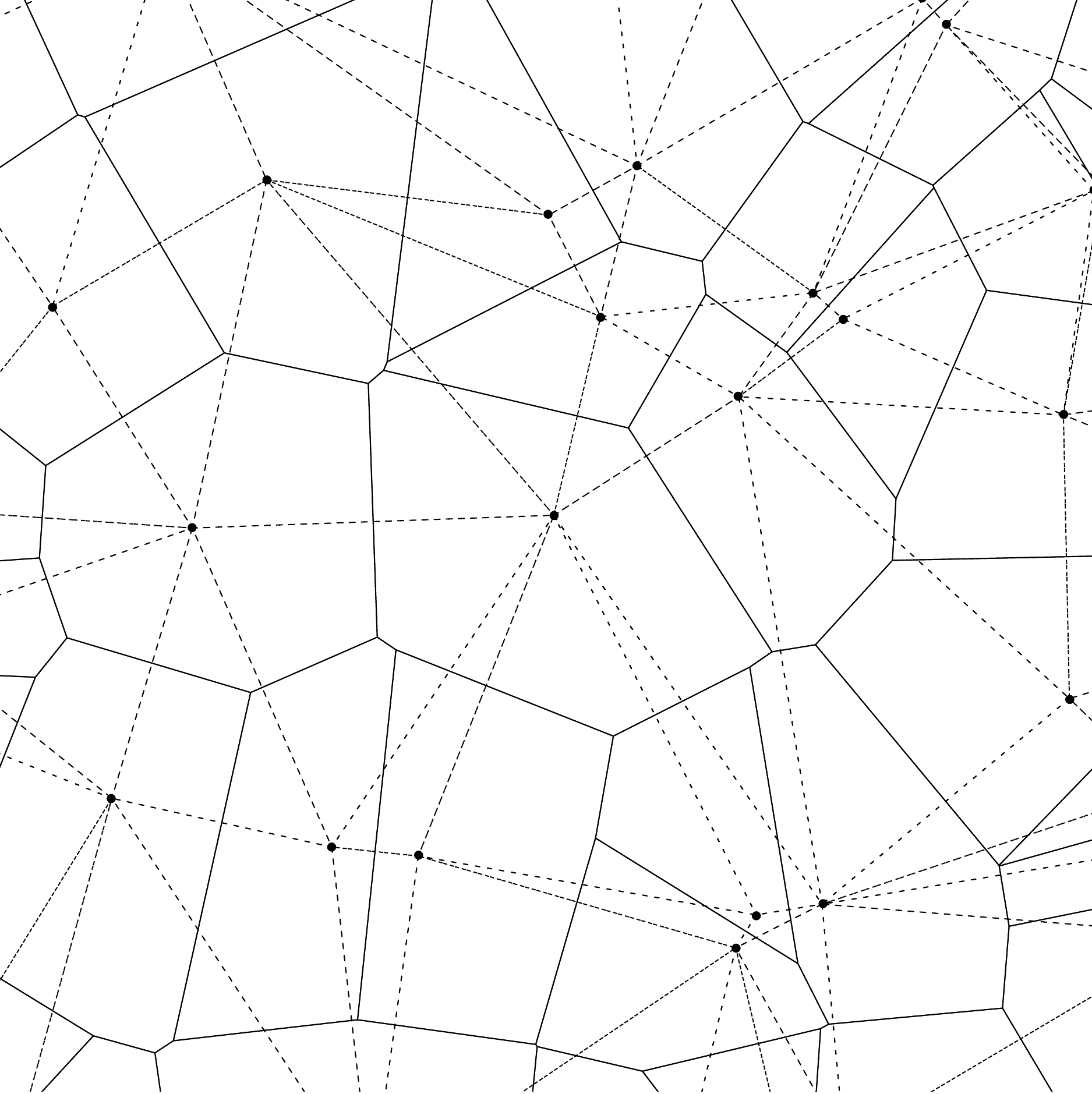} 
   \caption{The Delaunay triangulation (dotted lines) superposed on the Voronoi diagram (solid lines), its dual graph, 
   for a set of Poisson-distributed generating points.}
   \label{fig:VoronoiDelaunay}
\end{figure}

	The dual to the Voronoi diagram is interesting in its own right. Known as the Delaunay triangulation \cite{Delaunay34} (see Fig.\ \ref{fig:Delaunay}), it can be defined independently of the Voronoi diagram for the same set of points $P$ on $\mathbb R^2$: it is simply the set of all possible triangles formed from triples chosen out of $P$ whose circumscribed circles do not contain any other members of $P$ (Fig.\ \ref{fig:DelaunayCircumcircles}). The Delaunay triangulation and the Voronoi diagram for the same set of points can be seen in Fig.\ \ref{fig:VoronoiDelaunay}. Note that while the members of $P$ are sites in the Delaunay triangulation, they are not sites in the Voronoi network, whose sites are the vertices of the polygons; also note that the edges of the Voronoi diagram lie along the perpendicular bisectors of the edges of the Delaunay triangulation --- however, the edges of the Voronoi diagram do not always intersect the edges of the Delaunay triangulation, as seen in Fig.\ \ref{fig:VoronoiDelaunay}.  The Delaunay triangulation represents the connectivity of the Voronoi {\it tessellation}  of the surface.

	There are many algorithms for constructing these networks. The fastest ones run in $O(n \log n)$ time for general distributions of points \cite{Fortune87,ShamosHoey75,GuibasKnuthSharir92,LeeSchachter80}, where $n$ is the number of generating sites, and this has been proven to be the optimal worst-case performance \cite{ShamosHoey75}.  For a Poisson distribution of points on the plane, there are many $O(n)$ expected-time algorithms \cite{Dwyer91,SuDrysdale95,Maus84,BentleyWeideYao80}.

	In addition to being theoretically interesting \cite{Hilhorst05,Hilhorst09,Hilhorst08,Lucarini08,deOliveira08}, both the Voronoi diagram and the Delaunay triangulation are widely used in modeling and analyzing physical systems. They have seen use in lattice field theory and gauge theories \cite{ChristFriedbergLee82}, analyzing molecular dynamics of glassy liquids \cite{StarrGlotzer02}, detecting galaxy clusters \cite{Ramella01}, modeling the atomic structure and folding of proteins \cite{Gerstein95, Poupon04}, modeling plant ecosystems and plant epidemiology \cite{Deussen98}, solving wireless signal routing problems \cite{Meguerdichian01,Bandyopadhyay03}, assisting with peer-to-peer (P2P) network construction \cite{Naor2007}, the finite-element method of solving differential equations \cite{Zienkiewicz05}, game theory \cite{Ahn04,Cheong04}, modeling fragmentation \cite{Kiang66}, and numerous other areas \cite{Okabe2000,Aurenhammer91}.
	
	Percolation theory is used to describe a wide variety of natural phenomena \cite{StaufferAharony94,Sahimi94}. In the nearly seventy years since the first papers on percolation appeared \cite{Flory41,BroadbentHammersley57}, it has become a paradigmatic example of a continuous phase transition. For a given network, finding the critical probability, $p_c$, at which the percolation transition occurs is a problem of particular interest. $p_c$ has been found analytically for certain 2D networks \cite{SykesEssam64,Wierman84,ZiffScullard06}; however, most networks remain analytically intractable. Numerical methods have been used to find $p_c$ for many such networks, e.g., \cite{StaufferAharony94,ZiffSuding97,LorenzZiff01,FengDengBlote08,QuintanillaTorquatoZiff00,Ballesteros99,Grassberger03}.

	In this paper we consider the percolation thresholds of the Voronoi and Delaunay networks for a Poisson distribution of generating points, as  represented in Figs.\ \ref{fig:Voronoi},  \ref{fig:Delaunay}, and \ref{fig:VoronoiDelaunay}.   There are four percolation thresholds related to these two networks: the site and bond percolation on each.
Being a fully triangulated network, the site percolation threshold of the Delaunay network is exactly $p_c^\mathit{site,Del} =\frac{1}{2}$ \cite{SykesEssam64,Kesten82,Menshikov86,Aizenman87}.  This result has recently been proven rigorously by Bollob\'as and Riordan \cite{BollobasRiordan06a}.  
Somewhat surprisingly, a search of the literature revealed no prior calculation of the site percolation threshold of the Voronoi network at all, despite the widespread use of such networks. A prediction for its value has recently been made by Neher, Mecke and Wagner \cite{NeherMeckeWagner08}; they use an empirical formula to predict $p_c^\mathit{site,Vor} = 0.7151$, but they too were unable to find any previous calculation of this value, either analytically or numerically.  (There are places in the literature (e.g. \cite{Okabe2000}) where the Voronoi ``site" threshold is listed as 1/2. This is true if the ``sites'' are taken to be the generating points from which the diagram is created, rather than the vertices of the diagram. Thus, this is actually the Voronoi tiling threshold, i.e.\ the percolation threshold of the Voronoi \textit{polygons}, which is in turn equivalent to the Delaunay site threshold, well known to be 1/2.) 

The bond thresholds of the Voronoi and Delaunay networks are complementary,
\begin{equation} 
p_c^\mathit{bond,Vor} = 1 - p_c^\mathit{bond,Del} \ ,
\label{eq:bond}
\end{equation}
because these networks are dual to one another \cite{Bollobas08}.
The first numerical measurement of the bond threshold for either network seems to be that of Jerauld et al.\ \cite{JerauldHatfieldScrivenDavis84}, who in 1984 found $p_c^\mathit{bond,Del}  =0.332$.  Shortly thereafter, Yuge and Hori \cite{YugeHori86} performed a renormalization group calculation which yielded $p_c^\mathit{bond,Del}  =0.3229$. In 1999, Hsu and Huang \cite{HsuHuang99} found $p_c^\mathit{bond,Del}  =0.3333(1)$ and $p_c^ \mathit{bond,Vor}= 0.6670(1)$ through Monte Carlo methods. (The numbers in parentheses represent the errors in the last digits.) These values led them to make the intriguing conjecture that the thresholds are exactly $1/3$ and $2/3$ respectively. There is, however, no known theoretical reason to believe that this conjecture is true. In order to test this conjecture, and to find the site percolation threshold of the Voronoi network, we have carried out extensive numerical simulations, as detailed below. In section \ref{sec:algorithm}, we describe our methods, and in section \ref{sec:results} we discuss our results and compare them to the thresholds of several related lattices, and also discuss the covering graph and further generalizations of the Voronoi system.  Conclusions are given in section \ref{sec:conclusions}.

\section{\label{sec:algorithm} Generating algorithms and analysis techniques}

\subsection{Delaunay/Voronoi generation algorithm}

In order to avoid edge effects in the networks when growing percolation clusters, and to make it possible to use more sites as seeds for those clusters (see subsection \ref{sec:clustergrowth}), we wished to create Voronoi and Delaunay networks with periodic boundary conditions. While popular fast algorithms for generating Voronoi and Delaunay networks exist, most notably the Quickhull algorithm \cite{Barber96}, these do not generally support periodic boundary conditions. We therefore created our own fairly straightforward algorithm for generating the desired networks. After coming up with it independently, we later found that it falls into the class of expected-linear-time algorithms known as Incremental Search \cite{SuDrysdale95}. The basic outline of the algorithm is as follows:
\begin{enumerate}
\item
Divide the region in which the generating points (vertices of the Delaunay triangulation) are located into squares of equal size (``bins'').
\item
Find a single Delaunay edge by picking a point at random and searching through its bin and neighboring bins to find the point which is its nearest neighbor. 
\item
Given a Delaunay edge and a ``side" to look on (immediately above or below the edge), determine the third point in the Delaunay triangle by looking at the radii of the circumcircles of the triangles formed by that edge with each point in its bin and all of the neighboring bins. 
\item
Look at the other Delaunay triangles that have been found in that bin and neighboring bins to make sure this new triangle is not a duplicate of one that has already been found. If it is not, find which of its neighbors have already been discovered and mark them as its neighbors, and vice versa.
\item
From the list of neighbors of the current triangle, figure out which of its edges are not already shared with neighbors (if any), and, if there are any unshared edges, whether the missing neighbor should be above or below the edge.
\item
Repeat steps 3-5 until there are no unprocessed edges left, at which point the Delaunay triangulation is finished. Because the neighbors of each triangle are known, this algorithm also yields an adjacency list of the sites on the Voronoi diagram (because the Voronoi diagram is dual to the Delaunay triangulation).
\end{enumerate}

\begin{figure}[ t] 
  \centering
   \includegraphics[width=0.5\textwidth]{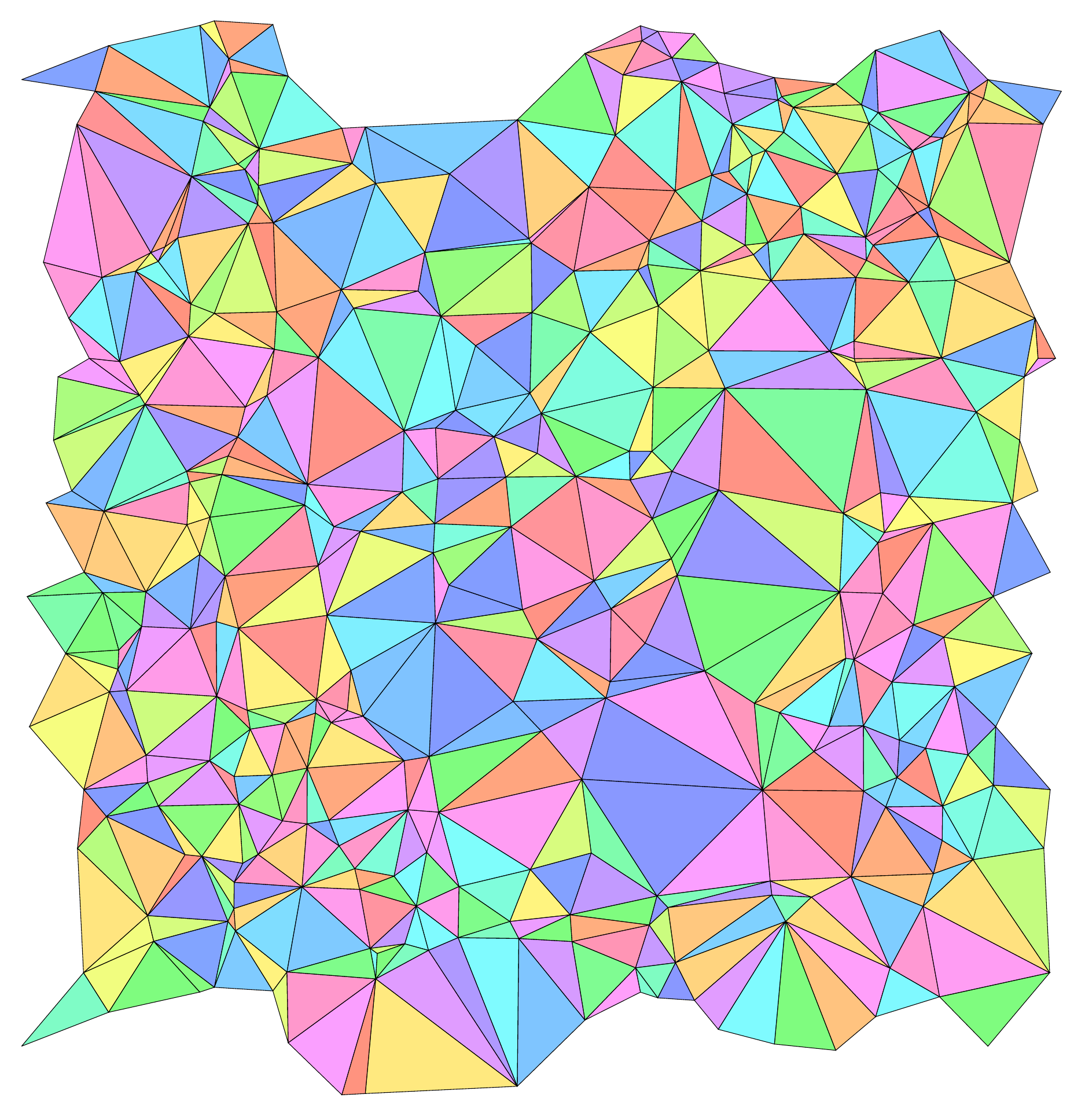}
   \caption{(Color online)  A Delaunay triangulation with periodic boundary conditions (i.e., on a torus), created from $n = 300$ generating points. Because the surface has periodic boundary conditions, there are exactly $2n = 600$ triangles here. Note the corresponding shapes of the outline on opposite edges, because this diagram has been ``unrolled" from a torus.}
   \label{fig:DelaunaySmall}
\end{figure}

This algorithm is significantly easier to implement with periodic boundary conditions, because every triangle is guaranteed to have exactly three neighbors. Furthermore, the imposition of periodic boundary conditions also gives the Delaunay and Voronoi networks a very useful property: there are always exactly twice as many Delaunay triangles (Voronoi sites) as there are generating points (vertices of the Delaunay network or polygons in the Voronoi network) for a given diagram. This is a consequence of the more general fact that that the number of faces (triangles) must be double the number of vertices (sites) for any fully triangulated network with doubly-periodic boundary conditions in two dimensions. This fact follows from Euler's formula and is proven in the appendix.   This simple relation makes it easier to spot certain kinds of errors in the code, because improperly written code is rather unlikely to consistently produce the proper number of sites for the given number of generating points. Using this algorithm, we generated thousands of Voronoi diagrams of 40,000 sites each. Fig.\  \ref{fig:DelaunaySmall} shows an example of a smaller Delaunay triangulation created with this algorithm.

\subsection{\label{sec:clustergrowth} Percolation cluster growth and finding $p_c$}

The Leath-type epidemic growth method \cite{Leath76} that we used involves growing a large number of percolation clusters in order to find $p_c$. For site percolation clusters, we start with a seed site somewhere on the network. Each of its neighbors is turned on with probability $p$ or off with probability $1-p$. Neighbors of active sites are then visited and the procedure is repeated for all their previously unvisited neighbors; the cluster either dies out naturally or is stopped by the program when it hits a cutoff size of 1000 sites. For bond percolation clusters, an analogous algorithm is used in which sites are simply never turned off, as it is the bonds between sites that are pertinent.
	
Due to the fact that our Voronoi diagrams are finite and are generated from a random Poisson distribution of points, each diagram yields a slightly different effective value of $p_c$; therefore, we had to generate many diagrams. This could have been very computationally expensive, but the choice of periodic boundary conditions helped here as well. Because there are no edges to the diagrams, we were able to place the seed point for a cluster at {\textit{any}} site on a diagram, rather than being limited to a small subset of sites near the center. This meant we were able to use many widely separated seed points to grow clusters on each diagram, which reduced the impact of each seed point's immediate neighborhood upon the value of $p_c$ obtained for each diagram. This, in turn, dramatically reduced the number of distinct diagrams we needed to obtain a particular level of precision. Specifically, we grew $8 \times10^5$ clusters of up to 1000 sites on each of 800 diagrams, for a total of $6.4 \times 10^8$ clusters grown at each value of $p$. We then repeated this process at each of various $p$ near $p_c$ to generate the plots in the next section. Finally, this process was done twice --- once for site percolation and once for bond percolation, both on the Voronoi network.
	
\begin{figure}[ t] 
   \centering
   \includegraphics[width=0.5\textwidth]{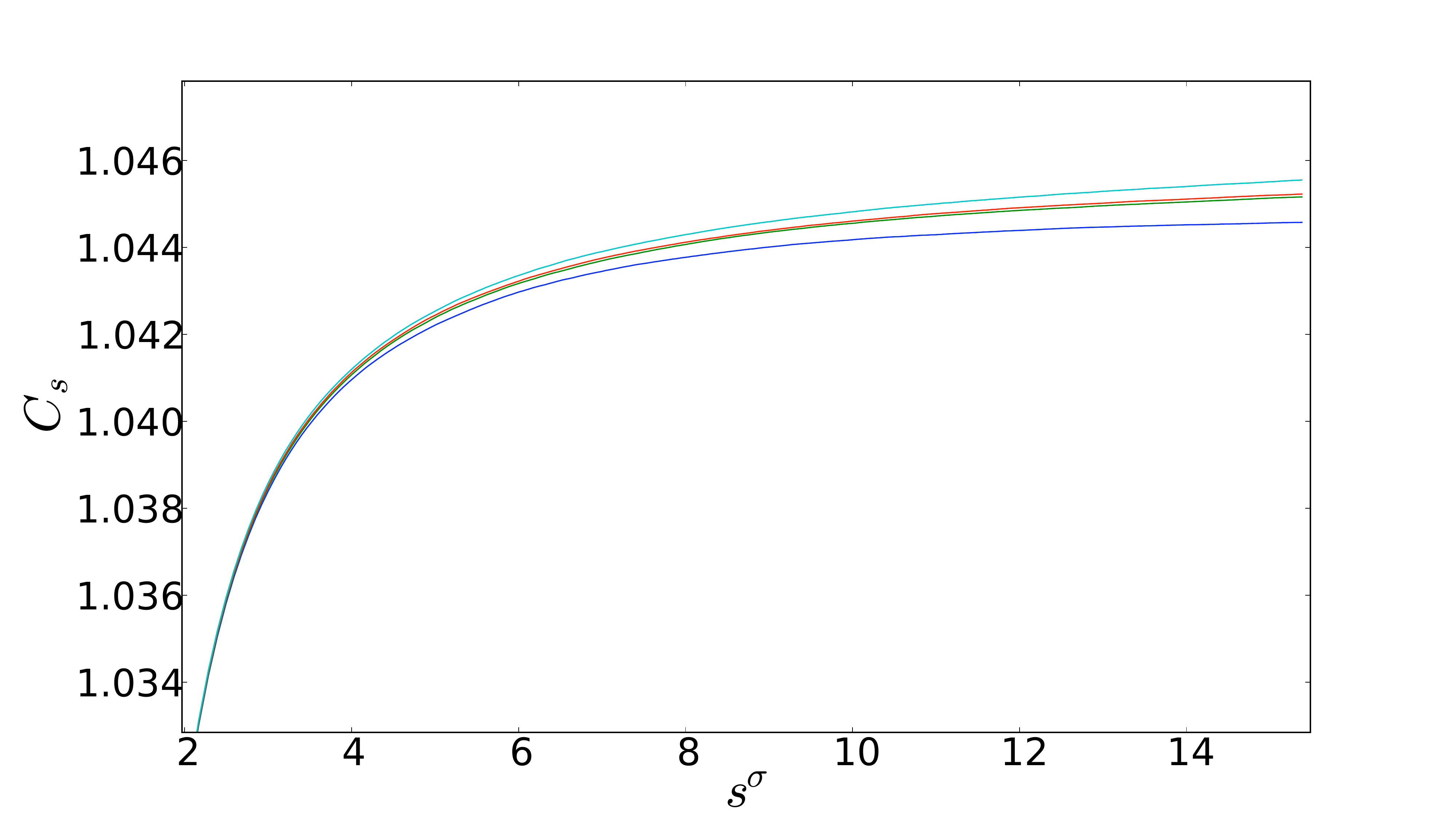} 
   \caption{(Color online) Epidemic site percolation cluster growth on the Voronoi diagram, for $p = 0.71407$, $0.71409$, $0.71411$, and $0.71413$, from bottom to top on the right.}
   \label{fig:PsPlot}
\end{figure}

\begin{figure}[ t] 
   \centering
   \includegraphics[width=0.5\textwidth]{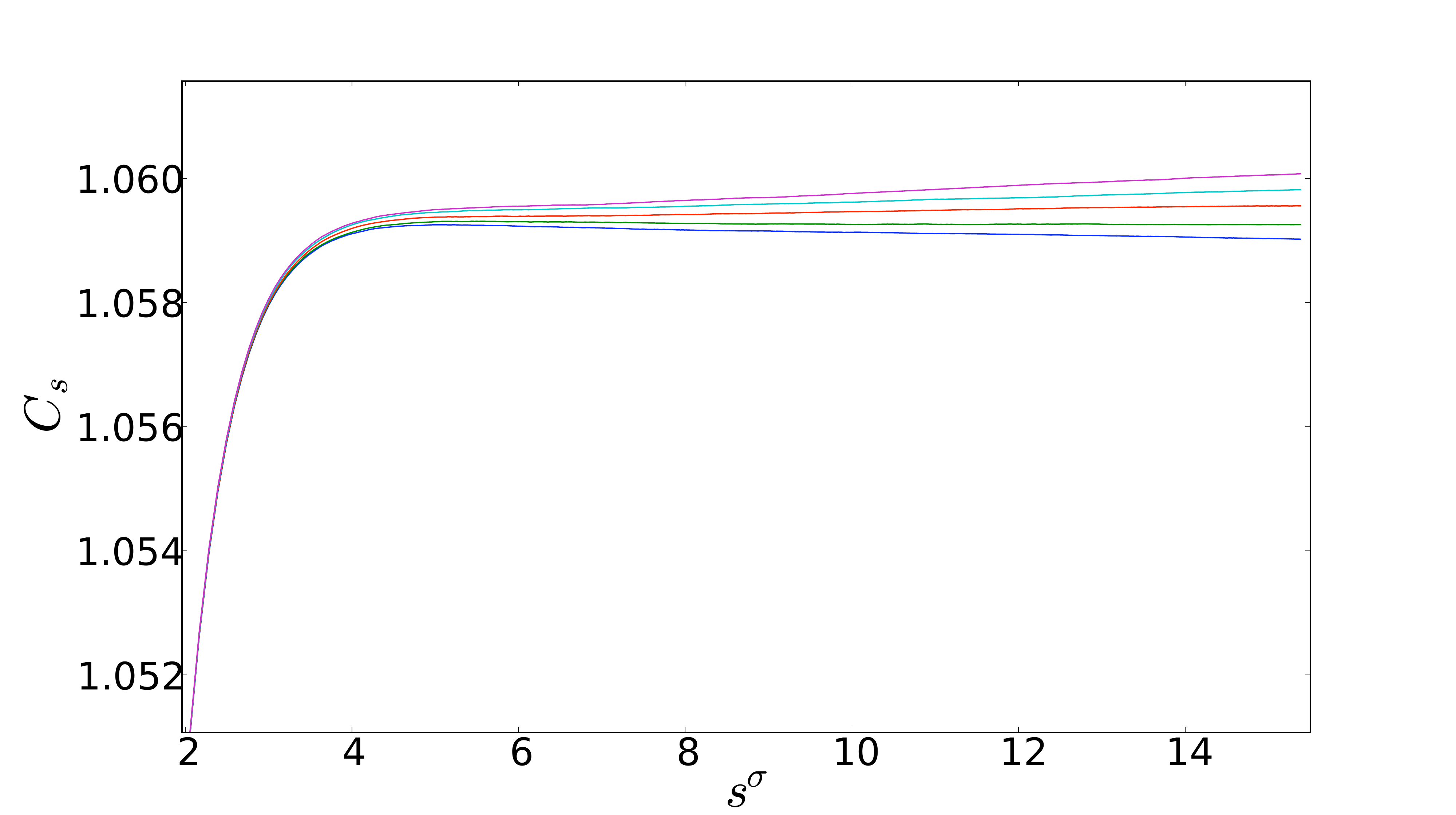} 
   \caption{(Color online) Epidemic bond percolation cluster growth on the Voronoi diagram, for $p = 0.66691$, $0.66693$, $0.66695$, $0.66697$, and $0.66699$, from bottom to top on the right. Note that these plots approach their linear asymptotes far more rapidly than those for site percolation clusters, as in Fig.\ \ref{fig:PsPlot}; also note the difference in the vertical scale between the two figures.}
   \label{fig:BondPsPlot}
\end{figure}

Because the percolation clusters are cut off before they can become large enough to wrap around the network, the clusters effectively see the diagram as infinite in size. Thus, their size distribution can be used to obtain an unbiased estimate of $P_{s}$, the probability that a percolation cluster will grow to be at least size $s$ (for $s \leq 1000$) on an infinite network. At the critical threshold $p_c$, $P_{s} \sim s^{2-\tau}$ as $s \rightarrow \infty$, where $\tau = 187/91$ for the two-dimensional percolation cluster universality class \cite{StaufferAharony94}. (It is expected that the critical exponents here are the same as for regular two-dimensional lattices.)  In the scaling region, where $s$ is large and $p - p_c$ is small such that $s^\sigma(p - p_c)$ is constant (with $\sigma = 36/91$), $P_s$ behaves as 
\begin{equation}
P_{s} \sim A s^{2-\tau} f(B (p - p_c)s^{\sigma}),
\end{equation}
where $A$ and $B$ are non-universal metric constants specific to the system being considered, and $f(x)$ is a universal scaling function analytic about $x = 0$.   If we operate close to $p_c$ such that 
$B (p - p_c)s^{\sigma} \ll 1$,
then we can make a Taylor-series expansion of $f(x)$ to find
\begin{equation} P_{s} \sim s^{2-\tau} (A + D (p - p_c) s^{\sigma} + \dots),
\label{ps_equation}
\end{equation}
where $D$ is another constant. Thus, plotting $C_s \equiv P_{s} \; s^{\tau-2} $ vs.\ $s^{\sigma}$ should yield a straight line at large $s$ when $p$ is near $p_c$, and that line will have a slope of zero when $p = p_c$.  Fig.\ \ref{fig:PsPlot} shows several such plots for site percolation clusters on the Voronoi network, and Fig.\ \ref{fig:BondPsPlot} shows several plots for bond percolation clusters on the same. $C_s$ does indeed approach a linear function for large $s$ in these plots, albeit far more quickly for bond percolation than for site percolation, with $p_c^\mathit{site,Vor}  \approx 0.7141$ and $p_c^\mathit{bond,Vor}  \approx 0.66693$.

Unfortunately, for smaller $s$ there are deviations in $C_s$ due to finite-size effects, and these are quite apparent for site percolation, even at the largest values of $s$ we were able to investigate.  Exactly at $p_c$, one expects
\begin{equation} 
P_{s} \sim s^{2-\tau}(A + E s^{-\Omega} + \dots)
\label{Ps_at_Pc}
\end{equation}
as $s \to \infty$, where $E$ is a constant and $\Omega \approx 0.6-0.8$ is the corrections-to-scaling exponent \cite{ZiffBabalievski99}. Similar deviations should occur when $p$ is close to $p_c$. In the case of site percolation on the Voronoi network, these finite-size effects make it difficult to determine when $C_s$ has a truly horizontal asymptote; thus, it is not possible to use the above method to find $p^{site,Vor}_c$ to much greater precision than four digits when $s \leq 1000$.

\begin{figure}[ht] 
   \centering
   \includegraphics[width=0.5\textwidth]{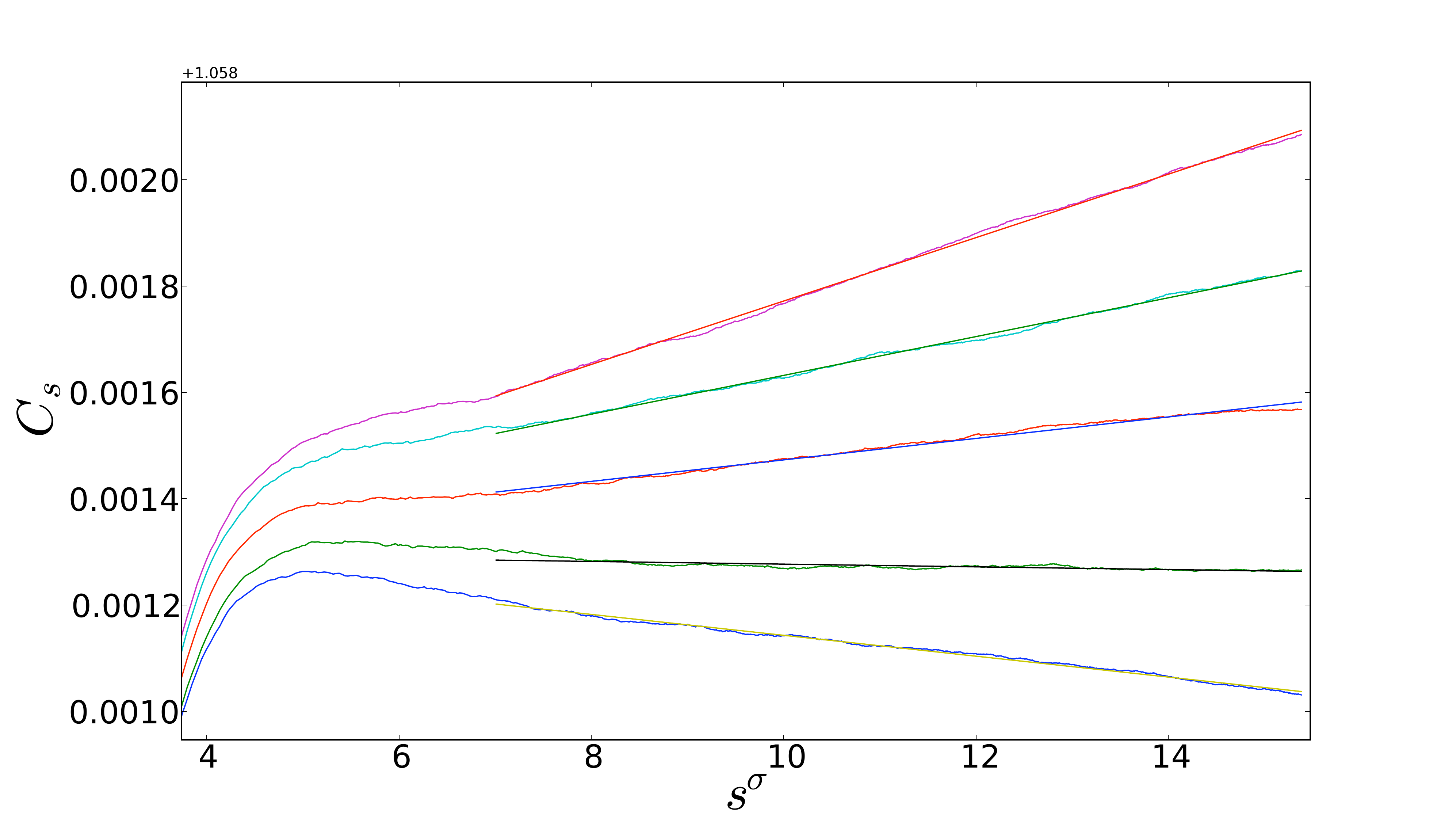} 
   \caption{(Color online) A zoomed-in portion of Fig.\ \ref{fig:BondPsPlot}. The deviations from the horizontal in the asymptotes for each curve can be seen more clearly here. (The $y$-axis values are given from a reference of $C_s = 1.058$). The least-squares linear fits for the curves are also on this plot.}
   \label{fig:ZoomBondPsPlot}
\end{figure}

The most straightforward way to solve this problem would be to grow larger site percolation clusters, using a larger system to insure that wrap-around does not occur. However, because of the computational time that would be required to do that, we instead used a more sensitive method to find $p_c$ that takes the finite-size corrections in (\ref{Ps_at_Pc}) into account.

\begin{figure}[ t] 
   \centering
   \includegraphics[width=0.5\textwidth]{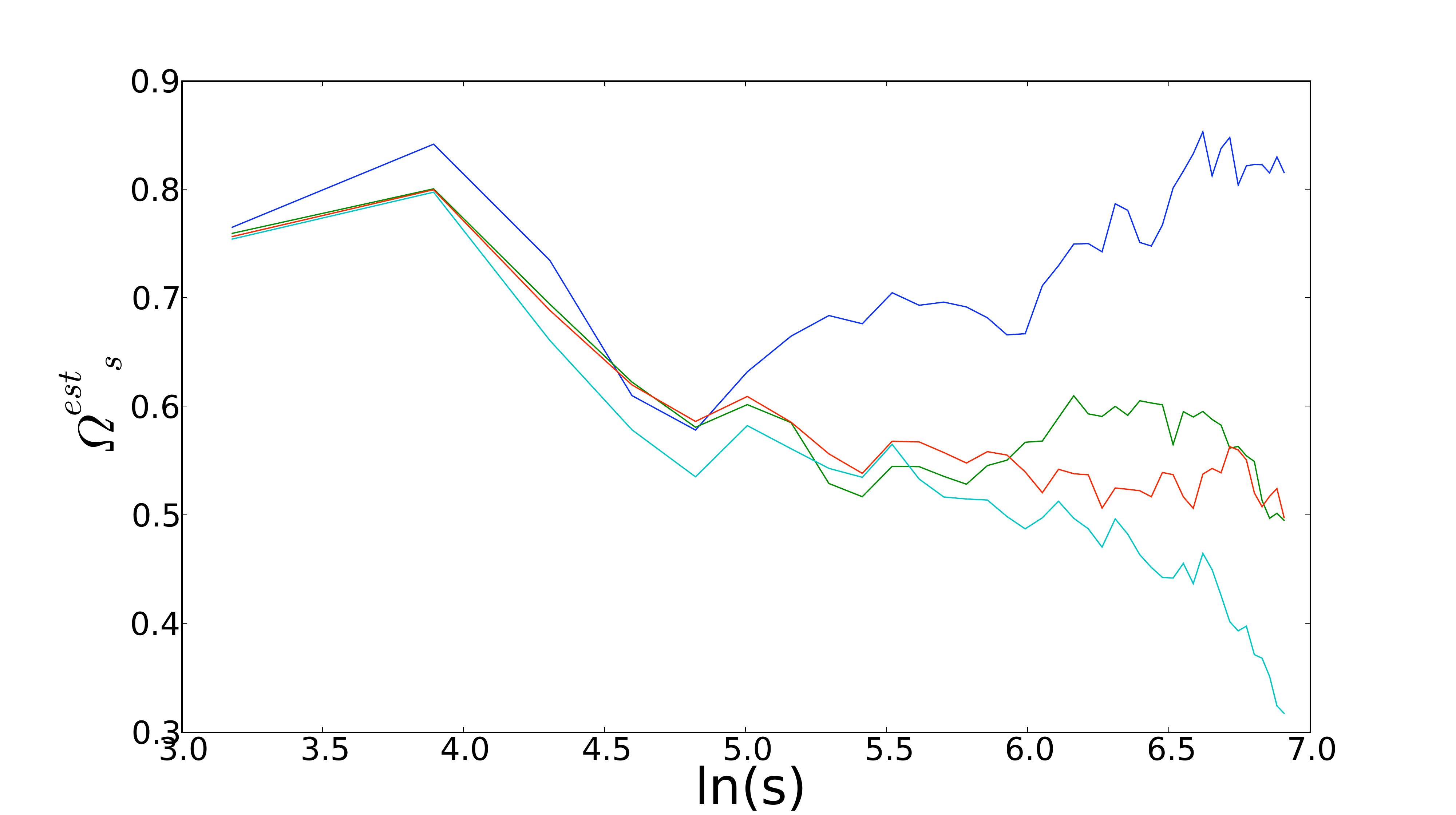} 
   \caption{(Color online)  $\Omega^\mathit{est}$ at $p = 0.71407,$ $0.71409,$ $0.71411,$ $0.71413$ (top to bottom on right) for site percolation on the Voronoi diagram.}
   \label{fig:OmegaPlot}
\end{figure}

Eq.\ (\ref{Ps_at_Pc}) implies that, at $p_c$, $C_s - C_{s/2} = E(1-2^{\Omega}) s^{-\Omega}$ to leading order. This means it's possible to estimate $\Omega$ directly from \cite{ZiffBabalievski99}
\begin{equation}
\Omega_s^\mathit{est} = -\log_2 \left( \frac{C_s - C_{s/2}}{C_{s/2} - C_{s/4}} \right).
\label{omega_equation}
\end{equation}
Thus, in the regime where $s$ is small enough that the finite-size effects of (\ref{Ps_at_Pc}) matter, yet large enough that higher-order corrections are unimportant, $\Omega_s^\mathit{est}$ should approach a constant $\Omega$ when $p = p_c$.  When  $p \neq p_c$, there will be deviations due to scaling. Plots of $\Omega_s^\mathit{est}$ vs.\ $\ln s$ for several values of $p$ can be seen in Fig.\ \ref{fig:OmegaPlot}; these yield the result for $p_c^\mathit{site,Vor}$ found in the following section. 

\section{\label{sec:results} Results and comparison with related lattices}

\subsection{$p_c$ for site and bond percolation on the Voronoi diagram}

Examining Fig.\ \ref{fig:OmegaPlot}, it can be seen that $\Omega_s^\mathit{est}$ approaches a constant for large $s$ for $p \approx 0.71409 - 0.71411$, 
and we conclude
\begin{equation} p_c^\mathit{site,Vor}  = 0.71410 \pm 0.00002\ ,
\label{pc_equation} \end{equation}
where the error bars are meant to indicate one standard deviation of error. This plot also gives us a rough value of 0.65 for $\Omega$ --- close to the value of $\Omega$ found for the Penrose rhomb quasi-lattice \cite{ZiffBabalievski99}.

\begin{figure}[ t] 
   \centering
   \includegraphics[width=0.5\textwidth]{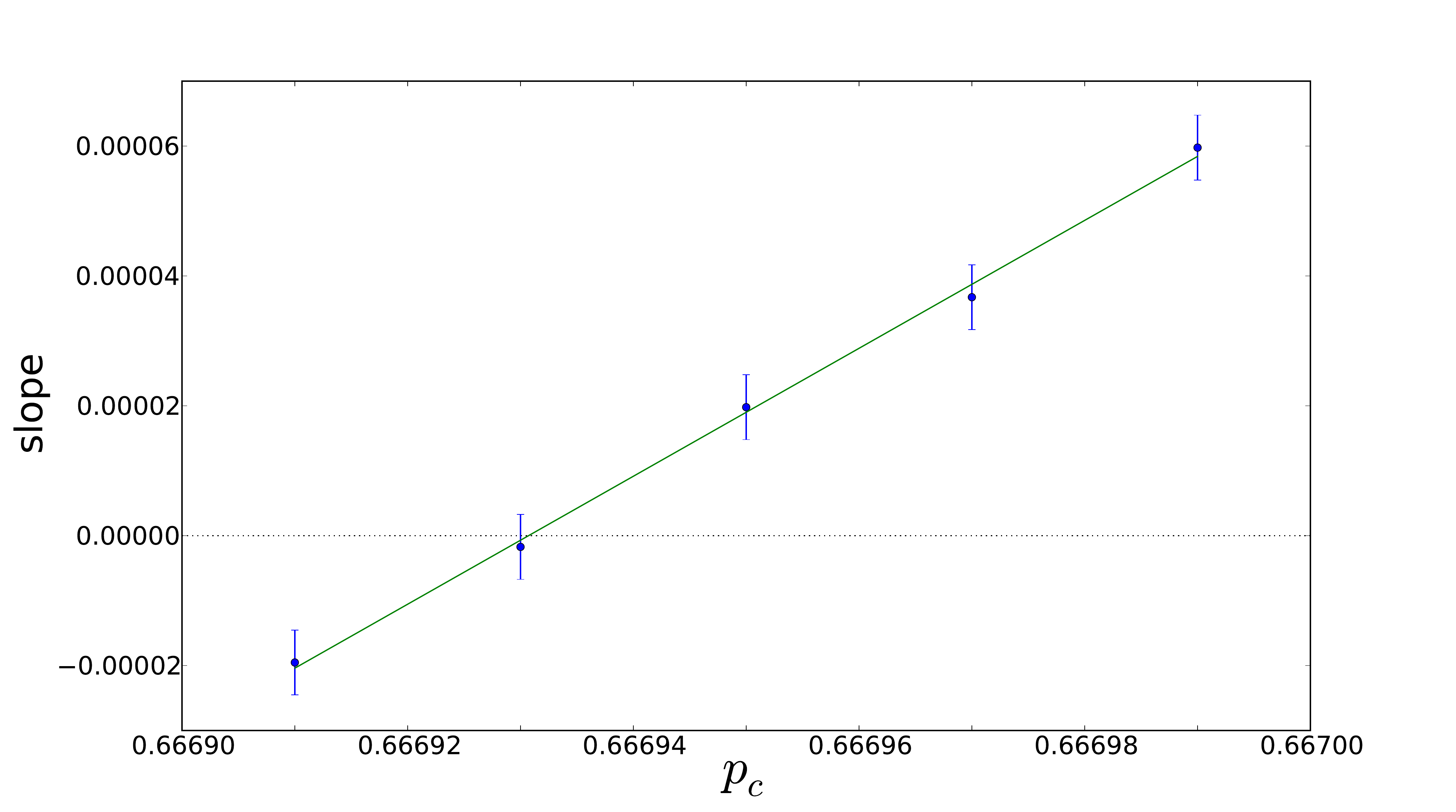} 
   \caption{Slopes of the lines fitted in Fig.\  \ref{fig:ZoomBondPsPlot} versus the values of $p$ used for each line, along with a best fit line. }
   \label{fig:BondSlopePlot}
\end{figure}

\begin{figure}[t] 
   \centering
   \includegraphics[width=0.5\textwidth]{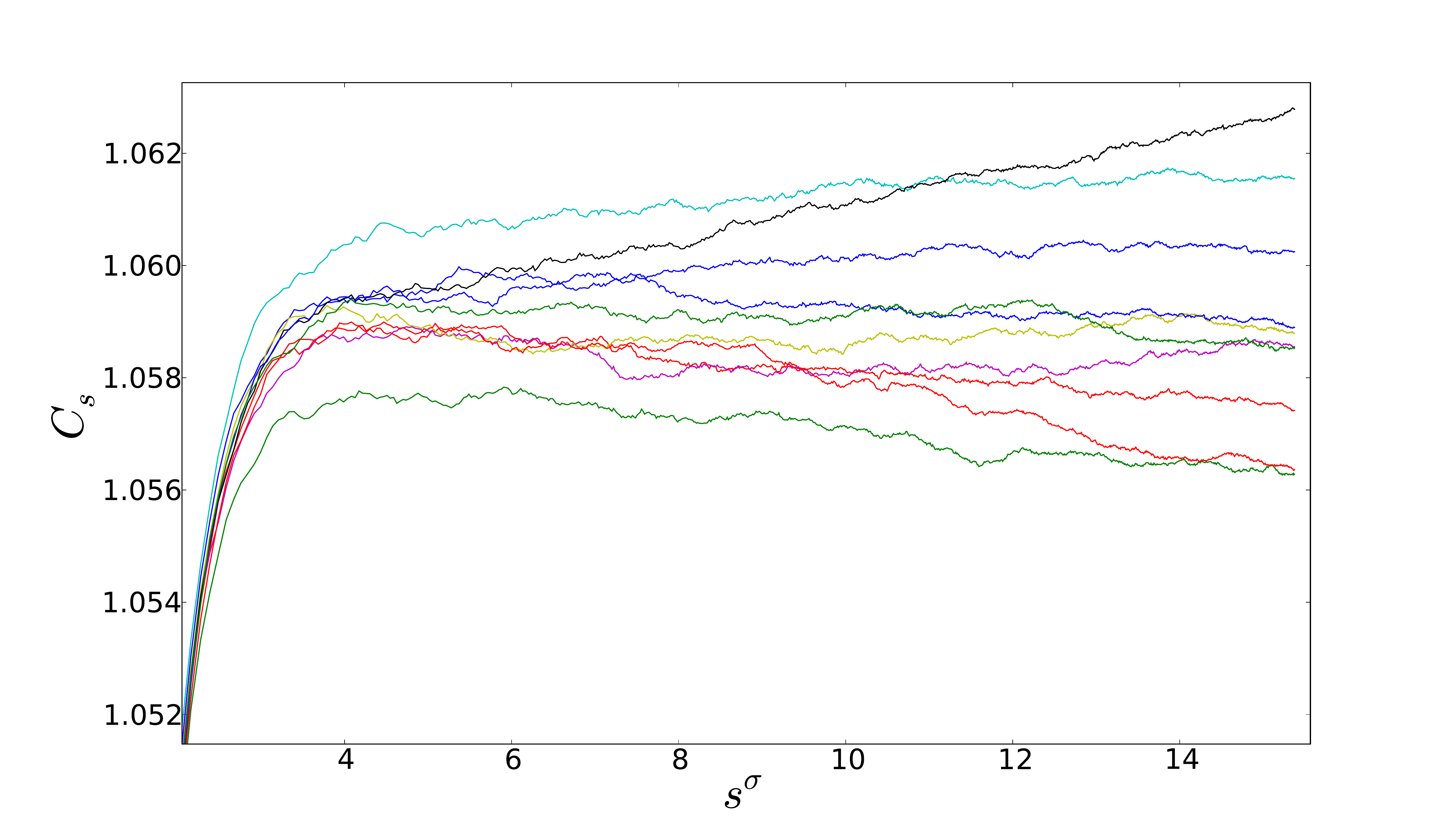} 
   \includegraphics[width=0.5\textwidth]{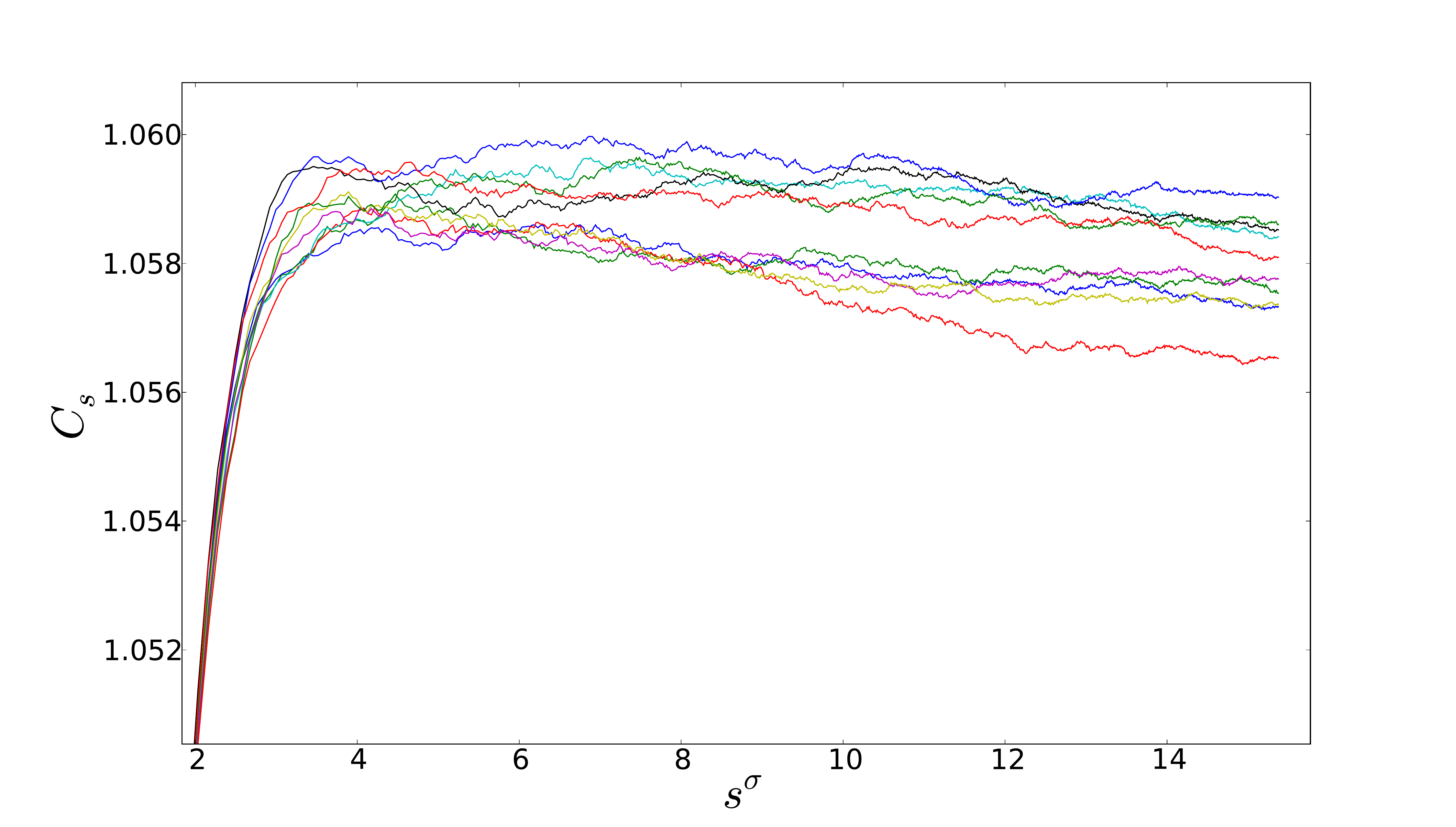} 
   \caption{(Color online)  Comparison of multiple-diagram and single-diagram bond percolation cluster growth on finite periodic Voronoi networks.  
   Each curve represents $8 \times 10^5$ clusters grown on 10 different diagrams (upper) or on 10 identical diagrams (lower), at $p  = 0.666931$.
   The mean value of the slopes are $-1.8 \times 10^{-5}$ (upper) and $-1.1 \times 10^{-4}$ (lower), and the standard deviations are $1.5 \times 10^{-4}$ (upper) and $4.7 \times 10^{-5}$ (lower).
   }
\label{fig:MultipleDiagrams}
\end{figure}

We used the method of plotting $C_s \equiv P_{s} \; s^{\tau-2} $ vs.\ $s^{\sigma}$, outlined in the previous section, to find the bond percolation threshold of the Voronoi diagram. Taking the results shown in Figs.\ \ref{fig:BondPsPlot} and \ref{fig:ZoomBondPsPlot}, we see immediately that  $p_c^ \mathit{bond,Vor} \approx 0.66693$. Because finite-size effects were not significant for bond percolation, we were able to find excellent least-squares linear fits to the asymptotic portions of the curves in Fig.\ \ref{fig:ZoomBondPsPlot}. By plotting the slopes of these lines against the values of $p$ used (see Fig.\ \ref{fig:BondSlopePlot}), we were able to solve for the value of $p$ that would yield a slope of zero; this should be $p_c$. This technique yielded a more accurate estimate:
\begin{equation} 
p_c^\mathit{bond,Vor}  = 0.666931 \pm 0.000005\ ,
\label{bondpc_equation} 
\end{equation}
which by (\ref{eq:bond}) implies $p_c^\mathit{bond,Del}  = 0.333069\pm 0.000005$.  We considered various contributions to the stated error.  First of all, it is unclear precisely where the linear regime begins in Fig. \ref{fig:ZoomBondPsPlot}, and this leads to some uncertainty in the slopes we measured from the best-fit lines.  Statistical effects of course are a source of error.  However, a somewhat larger source of uncertainty turned out to be the error involved in reusing the same diagram multiple times during cluster growth --- even with different seed points, there is a distinct likelihood that the same part of the non-uniform diagram will be sampled.  To estimate this error, we considered our usual runs of 800,000 samples on 10 different diagrams at  $p = 0.666931$ and looked at the variation in the curves of $C_s$ vs.\ $s^\sigma$ (Fig.\  \ref{fig:MultipleDiagrams}).  In contrast, we also looked at 10 runs of 800,000 samples each on the same diagram, to gauge the purely statistical error.  We found the errors in the previous case larger than in the latter.  Using the measured standard deviation $1.5 \times 10^{-4}$ and dividing by $\sqrt{800}$ for the $800$ runs we actually used in our simulations for each value of $p$, we estimate a final error of $\pm 0.000005$ in the slopes of $C_s$, as indicated in the error bars of Fig.\ \ref{fig:BondSlopePlot}. 
Finally, because the slope of the fitted line in Fig.\ \ref{fig:BondSlopePlot} (which equals the coefficient $D$ in Eq.\ (\ref{ps_equation})) is nearly $1$, we estimate that the final error in $p_c$ is $\pm 0.000005$.  Note that the runs for the five values of $p$ were each done on 800 {\it different} diagrams, so there is no systematic error among the least-squares fit lines drawn in Fig.\ \ref{fig:ZoomBondPsPlot}.  Because of this, we believe our error bars are conservative.

The results for the thresholds are summarized in Table \ref{table:thresholds} and discussed further below.

\begin{table}
\caption{Results for percolation thresholds of Voronoi and Delaunay networks.  Numbers in parentheses represent errors in last digits.}
\label{tab:1}       
 \begin{tabular}{llll}
 \hline
 \hline 
 network   \qquad &  $z$ \qquad     & $p_c^\mathit{site}$   \qquad & $p_c^\mathit{bond}$ \\
\hline 
Voronoi  & 3 &0.71410(2)  & 0.666931(5)  \\
Delaunay & 6 (avg.) & 0.5 (exact) & 0.333069(2) \\
 \hline
\end{tabular}
\label{table:thresholds}
\end{table}


\subsection{Comparison with thresholds of related lattices}
 
The Voronoi network has a uniform coordination number $z$ equal to 3. In Table \ref{table:z3}, we compare the site and bond thresholds of the Voronoi diagram with several other lattices with  $z = 3$, listed in descending order of threshold values. In the Gr\"unbaum-Shepard notation, $(3^{a_3}, 4^{a_4}, \dots)$ describes a lattice with $a_3$ triangles, $a_4$ quadrilaterals, etc., per vertex.  The Archimedean lattices $(3, 12^2)$, $(4, 6, 12)$, and $(4, 8^2)$ are illustrated in \cite{Grunbaum,SudingZiff99,PercolationThresholdWiki}.  The martini lattice was introduced in \cite{Scullard06} and can be represented by  (3/4)(3, $9^2$) + (1/4) $(9^3)$. 

We also list in Table \ref{table:z3} for each lattice the generalized filling factor $f$, defined as \cite{SudingZiff99}
\begin{equation}
f = \pi \left[ \sum_{n \ge 3} a_n\cot \frac{\pi}{n} \right]^{-1}  ,
\label{eq:filling}          
\end{equation}
which generalizes Scher and Zallen's definition of $f$ for lattices not necessarily composed of regular polygons \cite{ScherZallen70}.  The $f$ has been shown to provide a good correlation to site percolation thresholds for a variety of lattices.  To calculate $f$ for the Voronoi network, we use  $b_3 = 0.0112400$, $b_4 = 0.1068454$, etc., from \cite{Hilhorst07}, where $b_n = 2 a_n / n$ is the fraction of $n$-sided polygons in the system, satisfying $\sum_n b_n = 1$ and $\overline n= \sum_n n b_n  = 6$ for $z = 3$.

In Table \ref{table:z3} also list the fluctuations in the number of the sides of the polygons for each lattice,
\begin{equation}
\frac{\mu}{\overline n^2} \equiv \frac{\overline{n^2} - \overline n^2 }{\overline n^2} \ ,
\end{equation}
which is equal to the fluctuations in the coordination number of the dual lattice.  It follows from Euler's formula that the average number of sides of the polygons, $\overline n= \sum_n n b_n$, in any 3-coordinated network is exactly six.  For the Voronoi diagram, $\mu \approx 1.7808116990$ is known exactly as an integral \cite{Brakke86,Hilhorst07,Finch03}.

\begin{table}
\caption{Thresholds of lattices with uniform coordination number $z = 3$, also showing the filling factor $f$ and polygon variance $\mu/\overline n^2$.  
$^a$Ref.\ \cite{SudingZiff99},
$^b$Ref.\ \cite{Parviainen07},
$^c$Ref.\ \cite{Scullard06},
$^d$Ref.\ \cite{Ziff06},
$^e$this work,
$^f$Ref.\ \cite{FengDengBlote08},
$^g$Ref.\ \cite{SykesEssam63},
$^*$exact.
}
  \centering
   \begin{tabular}{lllll}
 \hline
  \hline
lattice	& $\mu/{\overline n}^2$ & $f$ & $p_c^\mathit{site}$ & $p_c^{\mathit bond}$ \\
 \hline
$(3,12^2)$&	0.5	&0.39067&	$0.807901^{*a}$&	$0.740422^b$ \\
martini	&0.25	&0.47493	&$0.764826^{*c}$ &	$0.707107^{*d}$	 \\
$(4,6,12)$&	0.222222	&0.48601	&0.747806$^a$&	0.693734$^c$	 \\
$(4,8^2)$&	0.111111&	0.53901	&0.729724$^a$&	0.676802$^c$	 \\
Voronoi	&0.049468&	0.57351	&0.71410$^e$&	0.666931$^e$	 \\
honeycomb&	0.0&	0.60460&	$0.697040^{a,f}$	&$0.652704^{*g}$ \\
\hline 
\end{tabular}
\label{table:z3}
\end{table}

\begin{figure}[b] 
   \centering
   \includegraphics[width=0.5\textwidth]{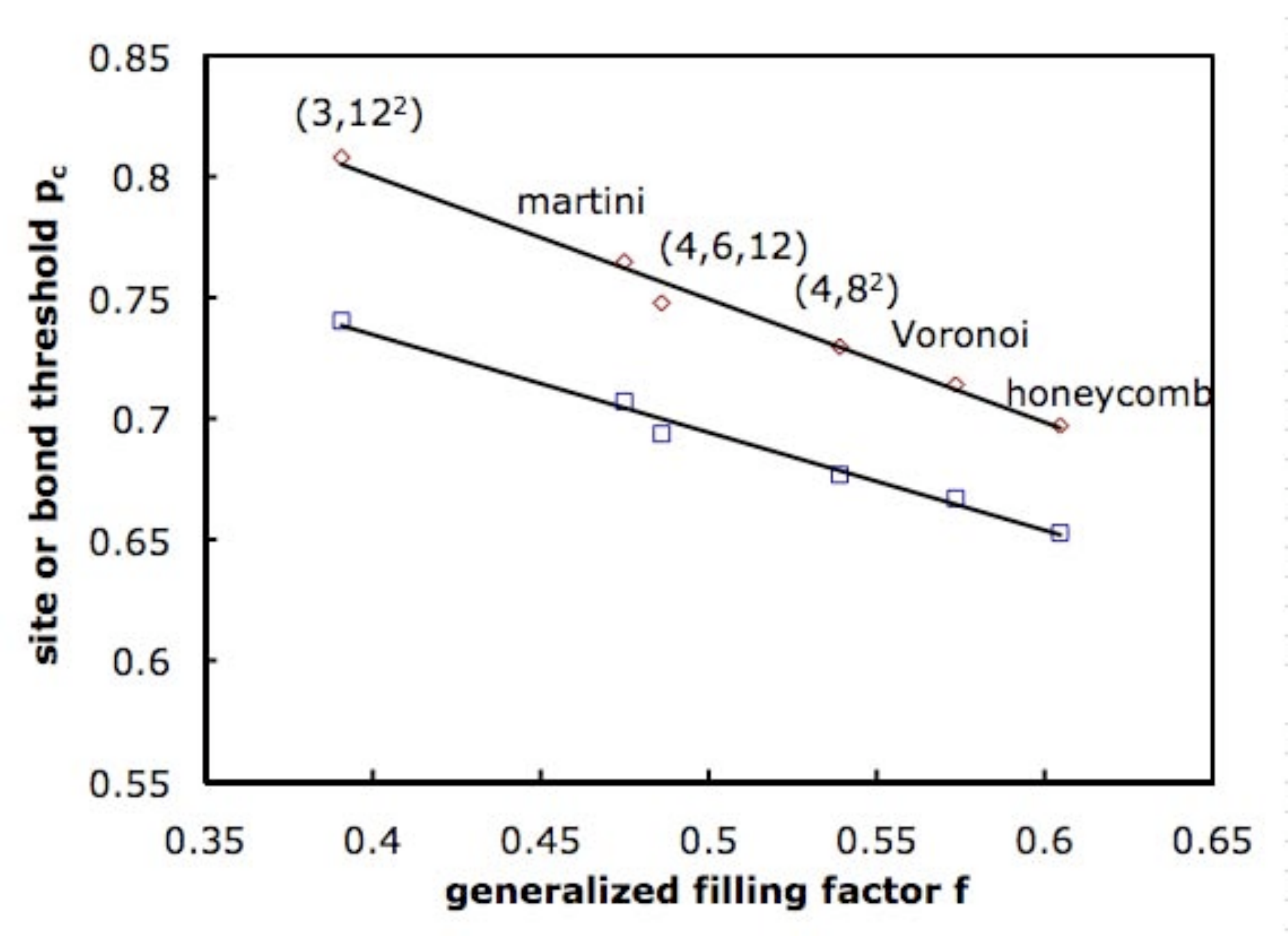} 
   \caption{Thresholds vs.\ generalized filling factor of Eq.\ (\ref{eq:filling}) for site (top) and bond (bottom) percolation for the systems of Table \ref{table:z3}.  The lines show least-squares fits to all of the data points.}
   \label{fig:filling}
\end{figure}

In Fig.\ \ref{fig:filling} we plot the thresholds given in Table  \ref{table:z3} as a function of $f$.  The thresholds fit well to a linear relation, as can be seen in the figure.
In general, for bond percolation, $f$ is not effective in correlating thresholds, which depend strongly upon the coordination number $z$.  However, for networks with fixed $z =3$, we find that the correlation of the bond thresholds with $f$ is quite good.

We can fit the linear behavior of $p_c(f)$ using just data from exact results,
with no numerical input.  For site percolation, we use the exactly known thresholds for the
 (3,12$^2$) and martini lattices, while for bond percolation we use the martini and honeycomb
 lattice results, and find:
 \begin{eqnarray}
p_c^ \mathit{site} &=& -0.5116 f + 1.0078 \ ,  \nonumber \\
p_c^ \mathit{bond} &=& -0.4195 f + 0.9063 \ . 
\end{eqnarray}
These equations imply for the Voronoi diagram (where $f = 0.57351$),  $p_c^ \mathit{site,Vor}  = 0.7143$ and $p_c^ \mathit{bond,Vor}  = 0.6657$, which are evidently excellent estimates.  Thus, the thresholds for the Voronoi diagram are consistent with other lattices with respect to the filling factor.  A similar plot of thresholds versus the fluctuations also shows consistent behavior between the Voronoi results and those for these other lattices.

\begin{figure}[ t] 
   \centering
   \includegraphics[width=0.45\textwidth]{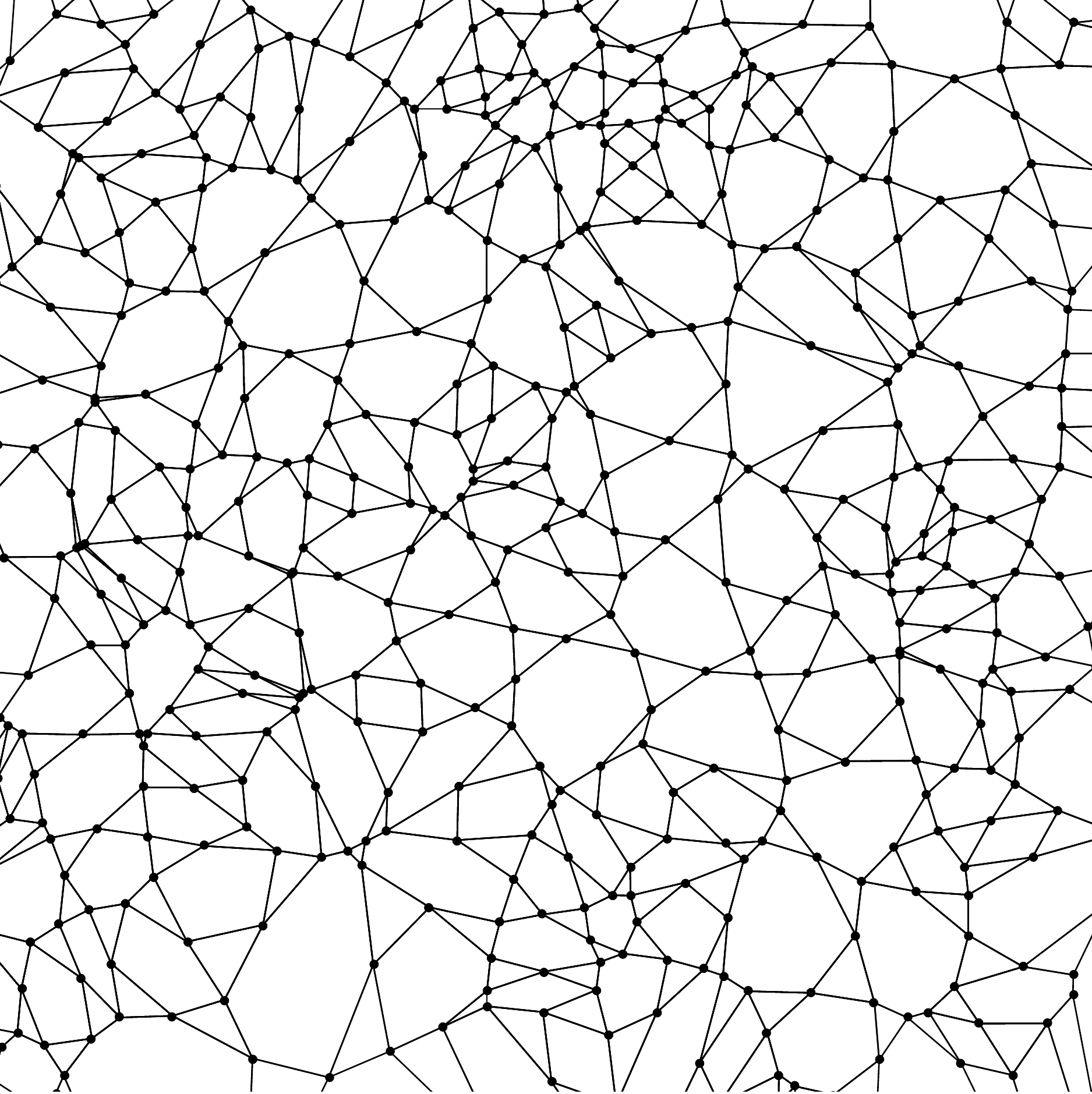} 
   \caption{Covering graph of a Voronoi network.}
   \label{fig:VoronoiCovering}
\end{figure}

Finally, the result for the bond threshold for the Voronoi network implies the site threshold for the Voronoi covering graph, shown in Fig.\  \ref{fig:VoronoiCovering}.  The covering graph (or line graph) for a given network is defined as the graph that connects the centers of the bonds together, and converts the bond percolation problem on that network to a site problem. Thus, $p_c^\mathit{site,Vor Cov} = p_c^\mathit{bond,Vor} \approx 0.666931.$ The covering graph is a kind of randomized kagom\'e diagram, consisting of triangles connected together.  Using similar arguments given in \cite{ZiffGu09} for generalized kagom\'e lattices, one can find an estimate for the bond threshold of the covering lattice, with the prediction $p_c^ \mathit{bond,Vor Cov} \approx 0.53618$, as well as an estimate for the site-bond threshold for the Voronoi diagram.  Details will be given elsewhere \cite{ZiffBeckerTBP}.

\section{\label{sec:conclusions}Conclusions}

We have determined the site percolation threshold for the Voronoi network, for the first time and to high precision, with the result $p_c^\mathit{site,Vor} = 0.71410(2)$.  We reiterate that this is not the well-known threshold (1/2) of the polygonal tiles of the Voronoi tessellation, which is equivalent to site percolation on the Delaunay triangulation, but rather the threshold for the 3-coordinated diagram of all the Voronoi polygons.  Our Monte-Carlo result is very close to the prediction 0.7151 of Neher, Mecke and Wagner \cite{NeherMeckeWagner08}, and confirms their empirical procedure based upon the Euler characteristic.

We also determined the bond threshold of the Voronoi network.  Our result $p_c^\mathit {bond,Del} = 0.333069(5)$  is consistent with Jerauld et al.'s result 0.332 \cite{JerauldHatfieldScrivenDavis84} and close to Hsu and Huang's value 0.3333(1) \cite{HsuHuang99}, but runs counter to the latter authors' conjecture that this threshold is exactly $1/3$.  It is interesting to note
that $1/3$ is the value predicted by the general (approximate) bond-threshold correlation $p_c \approx d/[(d-1)z]$ given by Vyssotsky et al.\ \cite{VyssotskyGordonFrischHammersley61} for $z = 6$ and dimension $d = 2$.

We made comparisons of our results with thresholds of other lattices with the same coordination number ($z=3$), and found that the Voronoi thresholds are what one would expect based upon correlations with the filling factor $f$ for both the site and bond problems.

Wierman has conjectured \cite{Wierman06} that $2 \sin \pi/18 \approx 0.3473$, the bond threshold of the regular triangular lattice, is the maximum possible bond threshold for any fully triangulated network, and that no other fully triangulated network has a bond threshold greater than or equal to that value. We indeed find that the bond threshold of the fully triangulated Delaunay network is consistent with this conjecture.  

For future work, it would also be interesting to look at thresholds for other random systems, such as Johnson-Mehl tessellations \cite{BollobasRiordan08} or the graph formed by the random distribution of lines in a plane \cite{Goudsmit45}.  Finding thresholds in Voronoi systems of higher dimensions is another interesting open problem.

\smallskip

\section*{ APPENDIX: Proof that $F = 2V$ for any fully triangulated network with doubly-periodic boundary conditions in two dimensions}
We take advantage of the Euler relation for polyhedra to prove the desired fact about fully triangulated networks. A network on a square surface with doubly-periodic boundary conditions is topologically equivalent to placing the network on the surface of a torus; this network, in turn, can be seen as a polyhedron on the surface of the torus. Thus, the Euler relation for polyhedra applies:
\begin{equation}
 V - E + F = \chi_\mathit{torus} = 0 
 \nonumber
\end{equation}
where $V$ is the number of vertices on the polyhedron, $E$ is the number of edges, $F$ the number of faces, and $\chi_\mathit{torus}$ the Euler characteristic for the 2-torus, which is zero. Because every face has exactly three edges (i.e., the network is fully triangulated), and every edge is shared by exactly two faces (the network has no boundary), we have $E = 3F/2$, and we can rewrite the Euler relation as follows:
\begin{equation} V - \frac{3F}{2} + F = V - \frac{F}{2} = 0
 \nonumber
\end{equation}
and thus $F = 2V$. QED.

\section*{ACKNOWLEDGMENTS}
This work was supported in part by the U. S. National Science Foundation Grant No.\ DMS-0553487.

\bibliographystyle{apsrev}
\bibliography{BeckerZiffPRE09resub}


\end{document}